# Electric Properties of Dirac Fermions Captured into 3D Nanoporous Graphene Networks

By *Yoichi Tanabe*[1†], *Yoshikazu Ito*[2†], *Katsuaki Sugawara*[2†], *Daisuke Hojo*[2†], *Mikito Koshino*[1†], *Takeshi Fujita*[2], *Tsutomu Aida*[3], *Xiandong Xu*[2], *Khuong Kim Huynh*[2], *Hidekazu Shimotani*[1], *Tadafumi Adschiri*[2], *Takashi Takahashi*[1,2], *Katsumi Tanigaki*[1,2], *Hideo Aoki*[4], and *Mingwei Chen*[2,5,6*]

Dr. Yoichi Tanabe, Prof. Mikito Koshino, Prof. Hidekazu Shimotani, Prof. Takashi Takahashi, Prof. Katsumi Tanigaki
Department of Physics, Graduate School of Science, Tohoku University,
Sendai, 980-8578, Japan

Dr. Yoshikazu Ito, Dr. Katsuaki Sugawara, Dr. Daisuke Hojo, Prof. Takeshi Fujita, Dr. Xiandong Xu, Dr. Khuong Kim Huynh, Prof. Tadafumi Adschiri, Prof. Takashi Takahashi, Prof. Katsumi Tanigaki, Prof. Mingwei Chen
WPI Advanced Institute for Materials Research, Tohoku University,
Sendai 980-8577, Japan
E-mail: mwchen@wpi-aimr.tohoku.ac.jp

Dr. Tsutomu Aida
New Industry Creation Hatchery Center (NICHe), Tohoku University,
Sendai 980-8579, Japan

Prof. Hideo Aoki
Department of Physics, University of Tokyo, Hongo,
Tokyo 113-0033, Japan

Prof. Mingwei Chen
CREST, Japan Science and Technology Agency,
Saitama 332-0012, Japan, and
Department of Materials Science and Engineering, Johns Hopkins University, Baltimore, MD 21214, USA

[†]These authors contributed equally to this work.

Keywords: (Graphene, 3D Nanoporous, Field effect transistor, Electric transport, CVD growth)

Graphene, as a promising material of post-silicon electronics, opens a new paradigm for the novel electronic properties and device applications [1-8]. The 2D carbon sheets characterized by massless Dirac fermions assure high-performance field effect transistors with a short channel and high mobility due to ultimate thickness and smooth surface [8]. 2D graphene transistors have shown 100 times higher carrier mobility ($2.5\times10^5$ cm$^2$V$^{-1}$S$^{-1}$) than that of Si [9]. On the other hand, the 2D feature of graphene makes it technically challenging to be integrated into 3D transistors with a sufficient processor capacity. Assembling 2D graphene into a 3D nanoarchitecture has been theoretically conceived as "graphite sponges" [10-12] for novel electronic properties, which are expected to realize high capacity and highly responsive graphene electric devices. Although there have been several reports on 3D graphene nanostructures [13-21], the characteristics of massless Dirac fermions cannot be well preserved in these materials for transistor applications because of a high density of crystal and topological defects produced by uncontrollable fabrication methods. Recently, high quality 3D nanoporous graphene with the well-preserved massless Dirac fermions has been successfully synthesized by chemical vapor deposition with a nanoporous metal template and supercritical drying [22]. The bicontinuous nanoporous graphene with a huge accessible surface area and retained 2D electronic properties is expected to provide a dramatically enhanced carrier injection in electric double layer transistors (EDLTs) [23]. In this study we experimentally realized a nanoporous graphene EDLT as illustrated in **Fig. 1(a)-(c)** and demonstrated for the first time that the 3D nanoporous graphene networks possess an ambipolor electronic nature of Dirac cones with an ultrahigh carrier mobility of 5000 – 7500 cm$^2$V$^{-1}$S$^{-1}$. Moreover, the 3D graphene networks with Dirac fermions turn out to exhibit a unique nonlinear Hall resistance in a wide range of the gate voltages. The high quality 3D nanoporous graphene EDLT may open a new field for utilizing Dirac fermions in 3D network structures for various fundamental and practical applications.

The nanoporous graphene materials were prepared with a nanoporous Ni based CVD method **[22]** (**See Supporting Information, Sections 1.1 and 1.2**). The bicontinuous nanoporous structure of the nanoporous Ni template **[24]** is well duplicated by atomically thin graphene **(Fig. 1(c))**, where the characteristic lengths of the porous structure can be tuned by controlling the CVD temperatures and durations **(Fig. 2(a, b))**. After completely dissolving the nanoporous Ni templates by chemical etching, the wet nanoporous graphene samples were processed by supercritical drying to avoid possible structural damages caused by capillary forces during drying (**See Supporting Information, Section 1.3**). The resultant free-standing nanoporous graphene samples with thicknesses of ~30 and ~120 μm are used in the present devices. The microstructure of the nanoporous graphene was characterized by scanning electron microscopy (SEM) and a smoothly interconnected graphene sheet in a 3D nanoporous structure can be seen **(Fig. 2(a, b))**. Transmission electron microscopy (TEM) further confirms the bicontinuous porous structure in which the graphene sheet and nanopore channels are continuously interconnected in 3D space **(Fig. 2(d,e))**. Different from other 3D graphene materials which are usually comprised of discrete graphene flakes, the 3D bicontinuous nanoporous graphene is constructed by a single-sheet graphene randomly distributed in 3D space without a seam and electrically weak interface. The CVD-grown nanoporous graphene has a high crystallinity as evidenced by the sharp diffraction spots in the selected area electron diffraction patterns **(Inset of Fig. 2(d,e))** from which detectable structural disorder cannot be found. The surface areas and pore sizes of the nanoporous graphene samples were measured by Brunauer–Emmett–Teller (BET) and Barrett-Joyner-Hallender (BJH) methods **(Fig. S1) [22, 25, 26]**. The sample grown at 800°C has the specific surface area of ~1220 $m^2g^{-1}$ with a pore size of ~ 228 nm, while the higher growth temperature of 900°C results in a coarsened nanoporous size of 1-2 μm and a decreased specific surface area of 964 $m^2g^{-1}$. For both samples, the porosity is larger than 99%. Such open commodious space can be easily filled with ion liquids to fully utilize the large surface area of graphene for high capacitance. Moreover, the diffusion kinetics of ions inside the porous samples is facilitated by the interconnected nanopore channels (**Fig. 1(c)**). The quality and electronic structure of the nanoporous graphene were characterized by Raman and photoemission spectroscopy (PES) (**See Supporting Information, Section 2.2, 2.3**). As shown in **Fig. 2(c)**, sharp 2D bands and high 2D/G intensity ratios in the Raman spectra taken by 514.5 nm laser suggest that the 3D nanoporous material is mainly composed of 1−2 layers of graphene **(Fig. 2 (d,e) and Table S1) [4, 27]**. For the 100–300 nm pore-sized sample, the appearance of a weak D band indicates a low density of topological defects in the 3D graphene network which are geometrically required to accommodate large curvature gradients in 3D nanoporosity **[22]**. The PES spectra of the nanoporous graphene materials exhibit a small shoulder at ~3 eV, while a large structure is seen at ~8 eV **(Fig. 2(f))**. These bands can be assigned respectively to carbon π and σ bands of 2D graphene with a small admixture from the *sp*$^3$ hybridization induced by the curvature gradient in 3D graphene **[22]**.

The nanoporous graphene EDLT devices with *N*, *N*-Diethyl-*N*-methyl-*N*-(2-methoxyethyl)ammonium bis(trifluoromethanesulfonylimide) as an ionic liquid gate were fabricated for electric transport studies **(Fig. 1(a) and (b), also See Supporting Information 1.4 and 2.4)**. The longitudinal conductance ($\sigma$), Hall coefficient ($R_H$) and capacitance ($C$) against the gate voltage ($V_G$) are shown in **Fig. 3 (a) - (b)**. The nanoporous graphene EDLT devices have ultrahigh conductance and capacitance ($\sigma \sim$ 0.05- 0.65 S, $C \sim$ 1 – 7 mF/cm$^2$), about 100 –1000 times larger than those achieved in 2D graphene EDLTs ($\sigma \sim$ 0.0005 – 0.006 S, $C \sim$ 0 – 0.005 mF/cm$^2$) **[28],** demonstrating that the electronic properties of the 3D porous nanoarchitecture can be successfully controlled by the ionic liquid gate **(See Supporting Infromation, Section 3.1 and Fig. S2)**. In comparison with previously reported 3D graphene network transistors ($\sigma \sim$0.002 – 0.008 S, $C \sim$ 0.00039 mF/cm$^2$) **[29]** and 2D graphene EDLTs, the significantly higher electric conductance and capacitance reveal that the large specific surface area of graphene can be effectively utilized by the 3D nanoporous graphene EDLTs. The unique 3D nanoarchitecture is thus expected to provide a significant advantage for device applications of graphene.

The electronic state of the nanoporous graphene against the gate voltage is displayed in **Fig. 3(c)**. With the increase of the gate voltage, the nanoporous graphene is tuned from the p-type to the n-type across the Dirac point. Accordingly, there are linear correlations among conductance, capacitance due to the accumulation of either electrons or holes, and the ambipolar carrier transport of Dirac fermions in the Hall conduction. Namely, as we change the gate voltage to accumulate electrons or holes, a linear increase of σ is observed in the nanoporous graphene with a conductance minimum at about $V_G$ = −0.6 V. This σ−$V_G$ characteristic is consistent with previous observations in graphene **[1-3, 28, 30, 31]**. As shown in **Fig. 3(b)**, the capacitance also linearly increases with the carrier accumulation of either electrons or holes at the turning point of about $V_G$ = −0.6 V. The coincidence in the turning (minimum) points for the conductance and the capacitance is consistent with the so-called quantum capacitance ($C_Q$), which is associated with the finite density of states of graphene **(See Supporting Information, Section 2.4) [31, 32]**. Moreover, the dependence of the Hall coefficient on the gate voltage (**Fig. 3a**) demonstrates that the charge neutrality point of the ambipolar carrier transport of Dirac fermions is at around $V_G$ = −0.6 V, consistent with previous reports of CVD-grown 2D graphene **[1-3, 31].**

**Figure 4 (a)** and **(b)** show the magnetic field ($B$) dependence of the longitudinal resistance and the Hall resistance ($R_{yx}$) of the nanoporous graphene at various values of $V_G$. The convex magnetoresistance against $B$ is observed at around $V_G$ = −0.6 V for higher $B$, while it exhibits a nearly linear behavior at the gate voltages away from $V_G$ = −0.6 V. The magnetoresistance shows a maximum at around $V_G = -0.6$ V, and decreases above and below this point. The gradient of $R_{yx}$ changes from positive to negative at around $V_G = -0.6$ V as well, which is consistent with the $R_H$-$V_G$ results in **Fig. 3 (a)**. The $B$ dependence of $R_{yx}$ exhibits a nonlinear curvature in the measured $V_G$ range, which has never been observed in 2D graphene before. If we turn to the curvature, $d^2R_{yx}/dB^2$ (**See Supporting Information, Fig. S4**), it shows a distinct peak at around $B = \pm 1$ T. It is worth noting that the Hall resistance curves of nanoporous graphene are symmetric, centered at $R_{yx} = 0$, with respect to the magnetic-field direction (**Fig. 4 (b)**). The high symmetry can simply rule out the possibility that the anomalous non-linearity in the Hall resistance is caused by ferromagnetic impurities, such as residual Ni **[33]**, in the nanoporous graphene.

To theoretically understand the electric transport phenomena of the nanoporous graphene, we considered the conductivity of 3D nanoporous graphene in a uniform magnetic field $B$. Here we assume that the normal of the local graphene surface is distributed isotropically so that the direction of $B$ is irrelevant and the conductivity contributed by surface elements is additive, although a sophisticated method such as the random network theory would give a more accurate picture. Based on the semiclassical transport theory (**See Supporting Information, Section 3.2**), the resistivity tensors are given as

$$\rho_{xx}(B) = \rho_0 \mu B \frac{\tan^{-1}\mu B}{(\tan^{-1}\mu B)^2 + [\log(1+(\mu B)^2)]^2/4} , \quad (1)$$

$$\rho_{xy}(B) = \rho_0 \mu B \frac{\frac{[\log(1+(\mu B)^2)]}{2}}{(\tan^{-1}\mu B)^2 + \frac{[\log(1+(\mu B)^2)]^2}{4}} , \quad (2)$$

where $\rho_0 = [(S_{\text{tot}}/V)\sigma_0^{2D}]^{-1}$, $\mu$ is the carrier mobility, $S_{\text{tot}}$ and $V$ are respectively the total graphene area and the total volume of the system, and $\sigma_0^{2D}$ is the conductivity of graphene at zero magnetic field. In **Fig. 4(c) - (d)**, we plot the magnetoresistivity $[(\rho_{xx} - \rho_0)/\rho_0]$ and the normalized Hall resistivity $(\rho_{xy}/\rho_0)$ against $\omega_c\tau = \mu B$, where $\omega_c$ is the cycrotron frequency and $\tau$ is the scattering time. We also plot the second derivative of the Hall resistivity, $d^2\rho_{xy}/dB^2$, in **Fig. 4 (d)**. The theoretical results well agree with the qualitative features of the experimental ones, specifically the characteristic kink structure in the second derivative of $\rho_{xy}$ which appears at $\mu B \sim 0.7$. We can then estimate the mobility of the nanoporous graphene from the kink position. It should be noted that the same analysis would not apply to usual continuum media with a single carrier component where the Hall resistance is always a linear in the magnetic field. Thus, the characteristic kink in $\rho_{xy}$ is expected to be a generic property peculiar to 3D labyrinth composed of a two-dimensional semimetal.

According to the semiclassical theory, μ$B$ should serve as a scaling parameter. We extract the amplitude of $B$ at the peak in $d^2R_{yx}/dB^2$ vs $B$ plot (**See Supporting Infromation, Fig. S4)**, and calculate μ using the condition of μ$B$ = 0.7 (**See Supporting Infromation, S6**). We re-plotted the dependence of normalized magnetoresitance [($R_{xx}$−$R_{xx}$(μ$B$ = 0))/ ($R_{xx}$(μ$B$ = 4) −$R_{xx}$ (μ$B$=0))], normalized Hall resistance [$R_{yx}$/$R_{yx}$(μ$B$ = 4)] and normalized second derivative of Hall resistance [(-$d^2R_{yx}$/$dB^2$)/ (-$d^2R_{yx}$(μ$B$ = 0.7)/$dB^2$)] against μ$B$ in **Fig. 4 (e,f)**. The values of $R_{xx}$, $R_{yx}$ and $d^2R_{yx}/dB^2$ are respectively normalized by $R_{xx}$, $R_{yx}$ at μ$B$ = 4 and $d^2R_{yx}/dB^2$ at μ$B$ = 0.7, since an accurate channel length is unclear. We can see that $R_{xx}$, $R_{yx}$ and $d^2R_{yx}/dB^2$ curves beautifully fall upon the scaling behavior with Eqs (1) and (2) at the gate voltages away from the Dirac point, i.e. $V_G$ = −2.1, −1.1, −0.1, 0.9 and 2.4 V, as shown in **Fig. 4 (e,f)**. When the gate voltage is close to the Dirac point with $V_G$ = −0.57, −0.60, and −0.63 V, $R_{yx}$ deviates from the scaling function while $R_{xx}$ can still be scaled for $V_G$ = −0.57, −0.60, and −0.63 V (**Supporting Information, Fig. S5**). However, [($R_{xx}$−$R_{xx}$(μ$B$ = 0))/ ($R_{xx}$(μ$B$ = 4) −$R_{xx}$ (μ$B$=0))] is slightly curved. The $V_G$ dependence of the carrier mobility is summarized in **Supporting Information, Fig. S6**, where μ lies in a range of 5000 – 7500 cm$^2$ V$^{-1}$S$^{-1}$, with a weak maximum around the charge neutral point. The high value of the mobility obtained here is comparable to those of high-quality 2D graphene prepared by CVD **[4]**.

By utilizing the EDLT devices, we demonstrated that the ambipolor electric transport as well as the quantum capacitance behavior of Dirac cones can be well preserved in the 3D nanoporous graphene networks. Note that a finite capacitance at about $V_G$ = −0.6 V may be influenced by the topological defects which are geometrically required to accommodate curvatures in 3D structure. The scattering of Dirac fermions by topological defects may give rise to a finite density of states for graphene at Dirac point, resulting in a finite capacitance [**30**]. As for the influence of the thicknesses of the 3D nanoporous graphene samples on transistor performances (See Fig. 3 (d) and **Supporting Information, Fig. S3 and Section 3.1**), the conductance of 100-300 nm pore sized samples with the thicknesses of 30 μm and 120 μm shows an nearly identical on/off ratio (3~4). For the 1 μm pore sized device, the on/off ratio of 120 μm thick nanoporous graphene is higher than that of 30 μm thick one, and close to these of 100 − 300 nm pore sized samples. Therefore, it seems that a stable device performance can be achieved when the thickness of the graphene sponges is about 100 times larger than the average pore sizes. Regarding the effect of pore sizes on the transistor performances of nanoporous graphene, the topological defects, which are geometrically required to coordinate curvature gradients in 3D nanoporosity, give rise to additional scattering in the samples with a smaller pore size. Therefore, the pore size dependence of conductance observed in this study is reasonable.

For the magnetotransport properties, both $R_{xx}$ and $R_{yx}$ curves follow the scaling behavior with Eqns (1) and (2) when the gate voltages are away from Dirac point. The deviation from the scaling around Dirac point is due to that the semiclassical argument becomes inapplicable for the massless Dirac fermions in the vicinity of Dirac point. For $V_G =$ −0.60 and −0.63 V, clear sign changes can be seen in the $R_{yx}$−$B$ curves, indicating that electrons and holes coexist in this $V_G$ region. For fermions on curved surfaces, the curvature of graphene can act as a potential due to a metric effect **[34]**, which affects the electronic band structure of graphene. If a similar curvature effect is exerted on Dirac fermions, this may also lead to the generation of both electrons and holes around the Dirac point. It has been suggested that a local curvature gives rise a mixing of π and σ bands, resulting in the enhancement of spin-orbit coupling in curved grapheme **[35]**. Experimentally, the mixture of π and σ bands has been observed in the nanoporous graphene by the angle integrated photoemission spectroscopy (PES) **[22].** However, the energy scale of the curvature induced spin-orbit coupling is only few Kelvin **[35]**, which is too small to influence the room-temperature magnetotransport. Therefore, the present magnetotransport data can be well described by the semiclassical model without the consideration of the curvature induced spin-orbit coupling.

The carrier mobility estimated from the scaling of the magnetotransport is as high as those in 2D graphene. Usually, assembling 2D graphene sheets into a 3D nanoarchitecture significantly degrades the carrier mobility because of electrically weak contact between the discrete graphene flakes **[13-21]**. Different from these 3D graphene, the present high carrier mobility is in line with the bicontinuous structure feature of the nanoporous graphene in which a single graphene sheet is smoothly arranged in the 3D space without electrically weak interfaces (**See Supporting Information Fig. S7**). Additionally, the interconnected open pore channels facilitate the realization of highly responsive graphene devices by utilizing the vast surface area of the 3D nanoporous structure.

To summarize, we have systematically investigated the properties of the Dirac fermions in 3D graphene networks with a unique bicontinuous porous nanoarchitecture. Utilizing the electric double layer transistors, the ambipolor electronic nature of Dirac fermions was unambiguously demonstrated in nanoporous graphene with the 3D morphology.

For the magnetotransport, the anomalous nonlinear Hall resistance was unveiled in a wide range of the gate voltage, which is elucidated to be an exceptional magnetotransport of Dirac fermions captured into the 3D graphene networks. Based on the scaling of magnetotransport from a semiclassical theory, the carrier mobility of the 3D graphene networks was estimated to be about 5000 – 7500 $cm^2V^{-1}S^{-1}$, which is comparable to those of flat 2D graphene sheets grown by CVD, which indicates that the 3D nanoporous graphene holds a great promise for applications in highly responsive electronic devices by utilizing the advantage of the vast surface area of the 3D nanoporous structure.

*Supporting Information*
Supporting Information is available online from Wiley Inter Science or from the author

*Acknowledgements*
This work was sponsored by JST-CREST "Phase Interface Science for Highly Efficient Energy Utilization"; the fusion research funds of "World Premier International (WPI) Research Center Initiative for Atoms, Molecules and Materials", MEXT, Japan;.Grant-in-Aid for Scientific Research on Innovative Areas "Science of Atomic Layers" (25107003, 26107504). KAKENHI 24740216, 24656028, 23224010, 24740193, 25107005, 15H05473, 25249108, 26289294.

Acknowledgements, general annotations, funding.((Supporting Information is available online from Wiley InterScience or from the author)).

Received: ((will be filled in by the editorial staff))
Revised: ((will be filled in by the editorial staff))
Published online: ((will be filled in by the editorial staff))

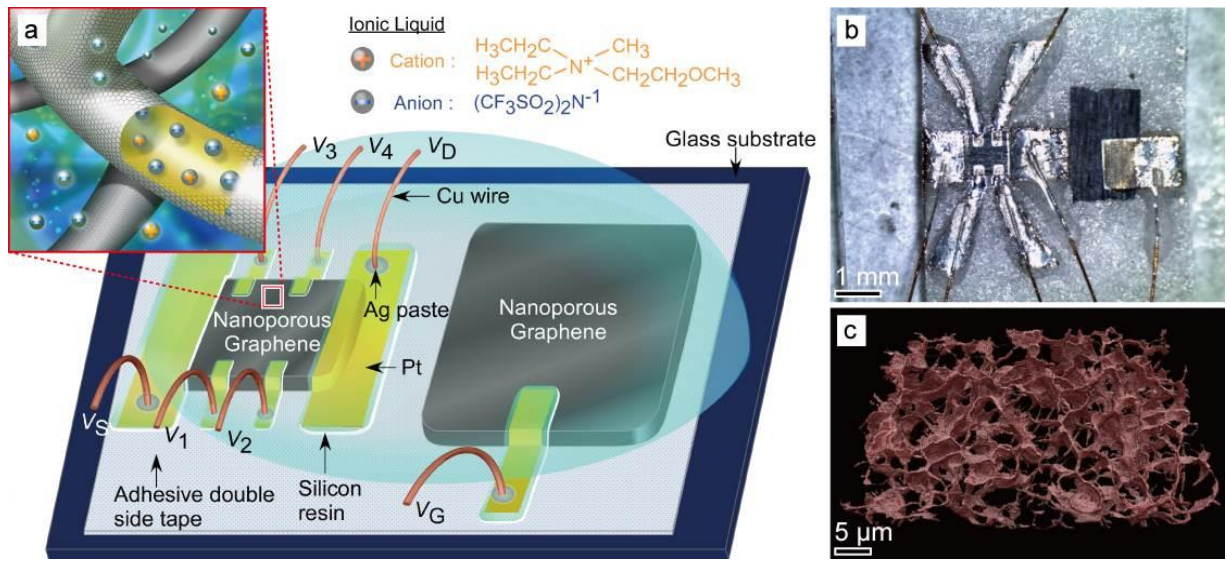

**Figure 1. 3D nanoporous graphene based electric double layer transistor (EDLT).** (**a**) A scheme of the nanoporous graphene EDLT. $V_S$, $V_D$, and $V_G$ are respectively the applied voltages for source, drain and gate electrode. $V_1$-$V_4$ are for the measurements of longitudinal resistance and Hall resistance. *N*, *N*-Diethyl-*N*-methyl-*N*-(2methoxyethyl)ammonium Bis(trifluoromethanesulfonyl)imide was employed as an ionic liquid gate. Inset schematically depicts that negative ions dominantly accumulate to the transistor channel side (with the gate electrode negatively charged), which induces a positively charged state on the transistor channel of the nanoporous graphene. The gate electrode was also prepared by the nanoporous graphene to exploit the large surface area. The present preparation processes of the material and the device structure permit the formation of EDL in the bicontinous nanopore structure, resulting in the tunable electronic state in the 3D transistor channel. (**b**) Zoom-in optical image of an actual device. (**c**) 3D structure of the nanoporous graphene with 1 μm pore size imaged by Focused Ion Beam tomography.

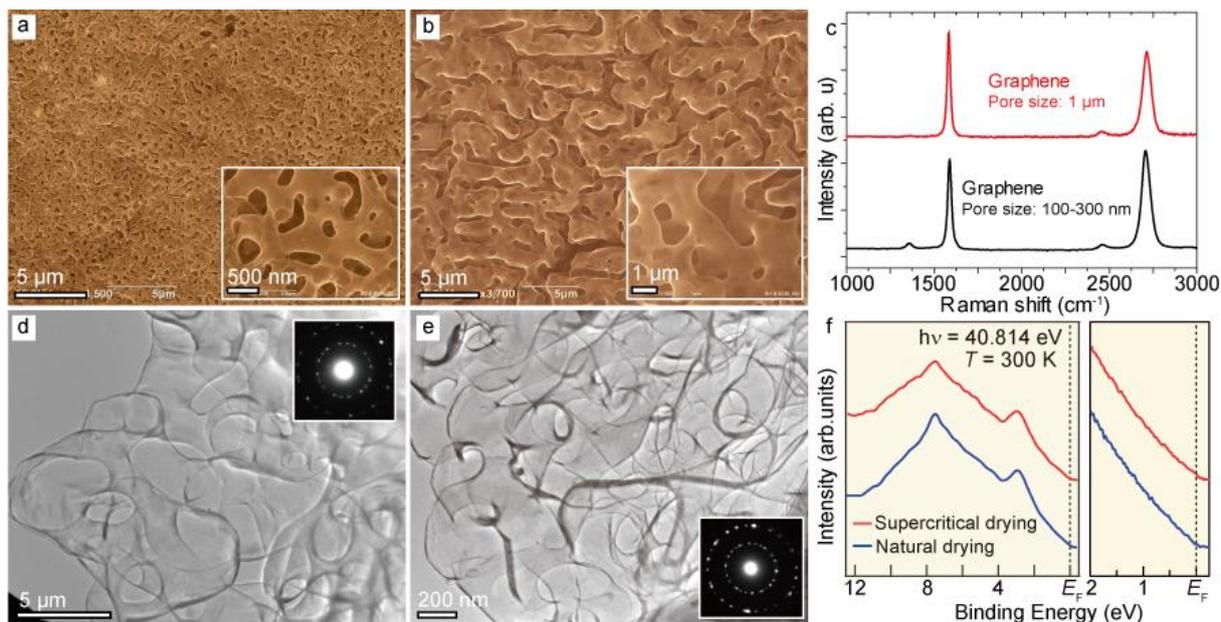

**Figure 2. Morphology and electronic density of states of 3D nanoporous graphene processed by supercritical drying.** Scanning electron beam microscopy (SEM) images of nanoporous graphene with (**a**) 100 – 300 nm pore size (inset: a magnified view); and (**b**) 1 μm pore size (inset: a magnified view). (**c**) Raman spectra of the nanoporous graphene samples with different pore sizes. (**d**) Typical low magnification TEM image and the selected area electron diffraction patterns of nanoporous graphene with 1.0 μm pores; and (**e**) nanoporous graphene with 100-300 nm pores. Transmission electron microscope images confirm smoothly bicontinuous porous structure in which either graphene or nanopore continuously distributes in the 3D space. Moreover, the selected area electron diffraction patterns (**Inset of (d) and (e)**) show sharp diffraction spots and multi-crystalline feature, demonstrating that the nanoporous graphene with a complex morphology retains high crystallinity while the interconnected graphene sheet is randomly distributed in 3D space. (**f**) Angle-integrated photoemission spectra of the nanoporous graphene with 100 – 300 nm pore size (with the supercritial drying treatment) measured by He IIa resonance line ($h\nu$ = 40.814 eV) at room temperature. For comparison, the spectra of nanoporous graphene with 100-300 nm prepared by the natural drying treatment are also plotted. Right panel is a magnified plot around the Fermi energy. Electronic structures do not differ significantly between the two samples, suggesting the enhanced EDLT performance by supercritial drying mainly from the conservation of damage-free 3D nanoporous structure of graphene.

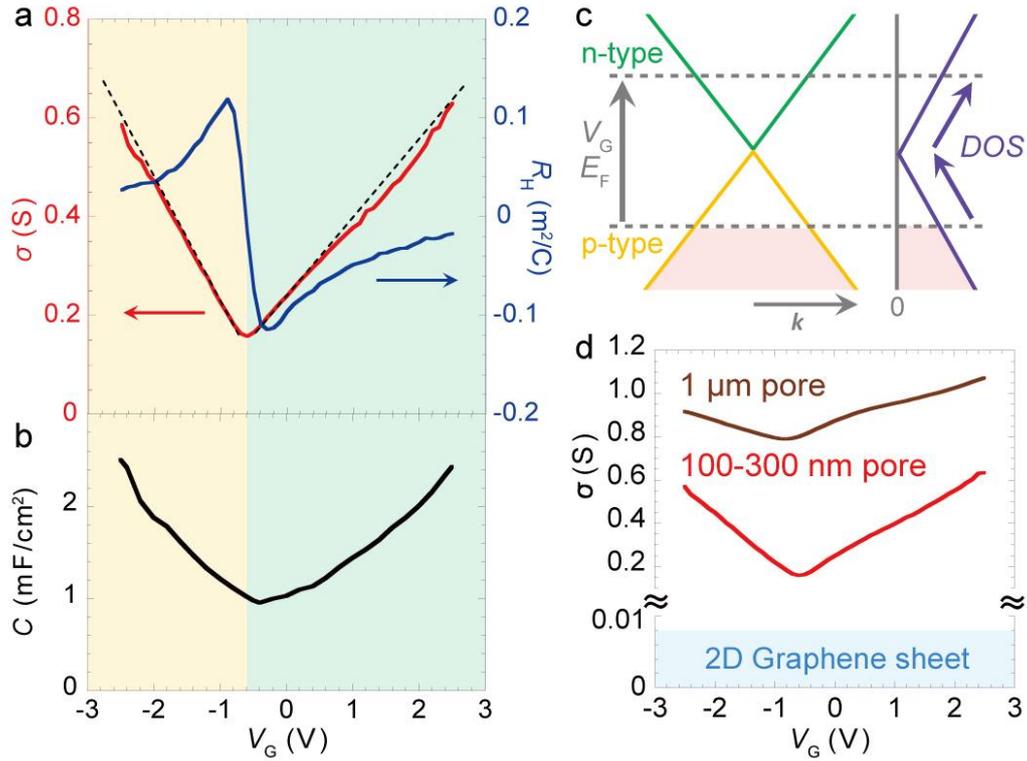

**Figure 3. Electric transport properties of 3D nanoporous graphene EDLT. (a)** and **(b)** The longitudinal conductance ($\sigma$), the Hall coefficient ($R_H$), and the capacitance ($C$) against the gate voltage ($V_G$) for the nanoporous graphene with 100-300 nm pores. $R_H$ is defined from the values of Hall resistance at magnetic fields of $\pm$ 9 T. Dashed lines are a guide for the eye. Both of $\sigma$ and $C$ exhibit pronounced V-shaped dips with a minimum around $V_G = -0.6$ V. For the EDLT device, the so-called quantum capacitance ($C_Q$) related to finite DOS **[31]** is dominant in the total capacitance ($C$) **[28, 32].** Thus, the nearly linear increase of $C$ with the accumulation of electrons or holes is consistent with the electronic structure of graphene around the Dirac point. Boundary between different shadings schematically indicates the turning point. **(c)** A schematic electronic structure of the nanoporous graphene with varying positions of $V_G$. **(d)** Dependence of $\sigma$ on $V_G$ for the nanoporous graphene with the pore sizes of 100-300 nm and 1 μm, respectively. For comparison, the shaded region represents the 2D graphene EDLT in the literature **[28]**. Namely, the conductance of the nanoporous graphene EDLT sits in a region of ~100-1000 times higher than those of 2D graphene EDLT.

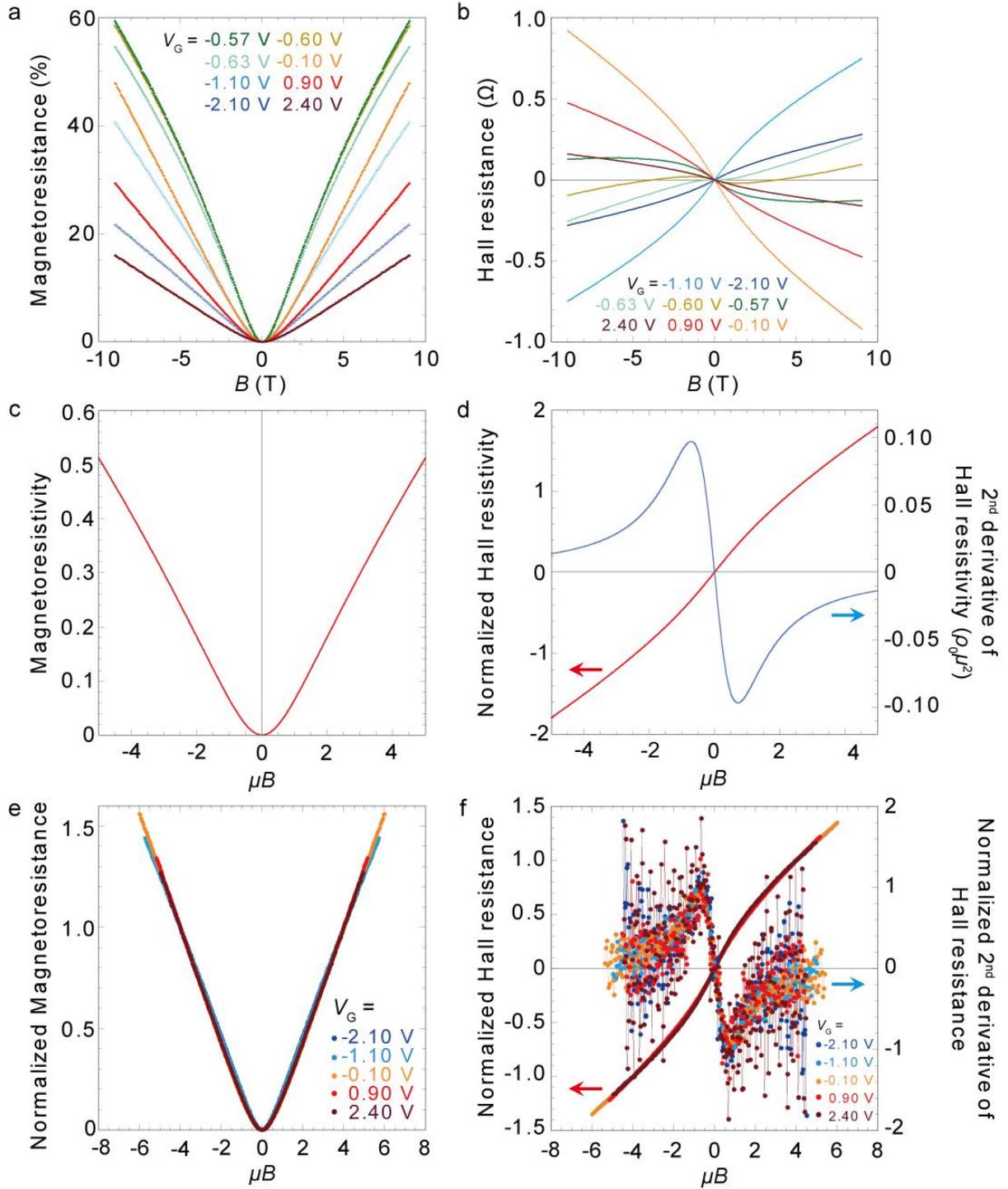

**Figure 4.** Magnetic field ($B$) dependence of (a) the magnetoresistance, and (b) the Hall resistance ($R_{yx}$). (c), (d) Theoretical magnetic field ($B$) dependence of magnetoresistivity [($\rho_{xx}(B)-\rho_0)/\rho_0$] and the Hall resistivity ($\rho_{xy}(B)/\rho_0$) for the nanoporous graphene in a semiclassical treatment. (e), (f) Dependence on $\mu B$, where $\mu$ is the carrier mobility estimated from the semiclassical transport theory for the nanoporous graphene as shown in Supporting Information Fig. S6. (e) displays the magnetoresistance [($R_{xx}-R_{xx}(\mu B=0))/(R_{xx}(\mu B = 4)-R_{xx}(\mu B=0))$]**,** while (f) displays the Hall resistance [$R_{yx}/R_{yx}(\mu B = 4)$] and the second derivative of Hall resistance [($-d^2R_{yx}/dB^2$)/ ($-d^2R_{yx}(\mu B = 0.7)/dB^2$)] for $V_G$ = -2.1, -1.1, -0.1, 0.9, and 2.4 V. Values of $R_{xx}$, $R_{yx}$ and $d^2R_{yx}/dB^2$ are respectively normalized by $R_{xx}$ and $R_{yx}$ at $\mu B = 4$ and $d^2R_{yx}/dB^2$ at $\mu B = 0.7$.

Supporting Information for

**Electric Properties of Dirac Fermions Captured into 3D Nanoporous Graphene Networks**

By *Yoichi Tanabe*[1†], *Yoshikazu Ito*[2†], *Katsuaki Sugawara*[2†], *Daisuke Hojo*[2†], *Mikito Koshino*[1†], *Takeshi Fujita*[2], *Tsutomu Aida*[3], *Xiandong Xu*[2], *Khuong Kim Huynh*[2], *Hidekazu Shimotani*[1], *Tadafumi Adschiri*[2], *Takashi Takahashi*[1,2], *Katsumi Tanigaki*[1,2], *Hideo Aoki*[4], and *Mingwei Chen*[2,5,6*]

Dr. Yoichi Tanabe, Prof. Mikito Koshino, Prof. Hidekazu Shimotani, Prof. Takashi Takahashi, Prof. Katsumi Tanigaki
Department of Physics, Graduate School of Science, Tohoku University,
Sendai, 980-8578, Japan

Dr. Yoshikazu Ito, Dr. Katsuaki Sugawara, Dr. Daisuke Hojo, Prof. Takeshi Fujita, Dr. Xiandong Xu, Dr. Khuong Kim Huynh, Prof. Tadafumi Adschiri, Prof. Takashi Takahashi, Prof. Katsumi Tanigaki, Prof. Mingwei Chen
WPI Advanced Institute for Materials Research, Tohoku University,
Sendai 980-8577, Japan
E-mail: mwchen@wpi-aimr.tohoku.ac.jp

Dr. Tsutomu Aida
New Industry Creation Hatchery Center (NICHe), Tohoku University,
Sendai 980-8579, Japan

Prof. Hideo Aoki
Department of Physics, University of Tokyo, Hongo,
Tokyo 113-0033, Japan

Prof. Mingwei Chen
CREST, Japan Science and Technology Agency,
Saitama 332-0012, Japan, and
Department of Materials Science and Engineering, Johns Hopkins University, Baltimore, MD 21214, USA

[†]These authors contributed equally to this work.

## 1. Materials and Methods
### 1.1 Fabrication of nanoporous Ni substrates

Ni$_{30}$Mn$_{70}$ (at. %) alloy ingots were prepared by melting pure Ni and Mn (purity >99.9 at.%) using an Ar-protected arc melting furnace. After annealing at 900°C for 24 hours for microstructure and composition homogenization, the ingots were cold-rolled to thin sheets with a thickness of ~50 μm. Nanoporous Ni was prepared by chemical dealloying in a 1.0 M (NH$_4$)$_2$SO$_4$ aqueous solution at 50°C[1]. After dealloying, the samples were rinsed thoroughly with water and ethanol and dried in vacuum.

### 1.2 Synthesis of nanoporous graphene by CVD

Nanoporous Ni substrates loaded in a quartz tube (φ26×φ22×250 mm) were inserted into a quartz tube (φ30×φ27×1000 mm) furnace and annealed at 800 or 900°C under flowing gas mixed by 2500 sccm Ar and 100 sccm H$_2$ for 3 min for 100 − 300 nm pores or 28 min for 1 μm pore sizes, respectively. After the reduction pre-treatment, benzene (0.5 mbar, 99.8%, anhydrous) [1, 2] was introduced with the gas flow of Ar (2500 sccm) and H$_2$ (100 sccm) at 800 or 900°C for 2 min graphene growth. Subsequently, the furnace was opened for quick cooling of the inner quartz tube. The nanoporous Ni substrates were dissolved by 1.0 M HCl solution and then transferred into 2.0 M HCl solution to completely removed residual Ni and Mn. The samples were repeatedly washed with water and isopropanol (IPA) and kept in isopropanol.

### 1.3 Procedure of supercritical drying

To prevent collapse and damage of the 3D nanoporous structure caused by capillary forces of water during drying, the 3D nanoporous graphene samples immersed in IPA were dried by using supercritical CO$_2$ (scCO$_2$) to substitute IPA by CO$_2$ with minimized capillary forces. The nanoporous graphene samples ware firstly transferred to a bottle (volume: 5 mL) filled with IPA (400 μL) and then the bottle was placed in an 80 mL pressure-resistant container (TAIATSU techno Corp). After removing air inside the container with CO$_2$ purging, the pressure of the container gradually increased to 15 MPa by introducing liquid CO$_2$ (5 MPa, -4°C, density of 0.964 g/mL)[3] at a flow rate of 20 mL/min ( ~19 g/min) using a high-pressure plunger pump (NIHON SEIMITSU KAGAKU Co. Ltd, NP-KX-540). The scCO$_2$ drying process was carried out at 70°C at a constant CO$_2$ flowing rate of 5 mL/min (~4.8 g/min) by forming a homogeneous phase of IPA and scCO$_2$ to minimize the capillary force. The pressure was maintained at 15 MPa during the drying procedure for 5 h. After drying completely, the temperature was set at 40°C and the container was gradually depressurized over 43 h from 15 MPa to atmospheric pressure by slowly pumping out CO$_2$ from the system.

### 1.4 Fabrications of nanoporous graphene electric double layer transistor (EDLT) devise

To utilize the vast surface area of the nanoporous graphene and the fast ion diffusion in the continuous open pore channels, the gate electrode was also prepared with the nanoporous graphene in the EDLT devices. Two pieces of nanoporous graphene were pasted on a glass substrate using the double faced tape. Both gate and channel electrodes were made by sputtering platinum. In order to prepare the electric contact between device and measurement system, Cu covered wire were attached on the electrodes with silver paste. Pt electrodes and Ag paste were covered by a silicone resin. An ionic liquid *N*, *N*-Diethyl-*N*-methyl-*N*-(2methoxyethyl)ammonium Bis(trifluoromethanesulfonyl)imide (DEME-TSFI), was employed to prepare the EDLT device, which has been proved to have stable EDLT performance in a large $V_G$ range with other carbon materials[4]. The DEME-TSFI was draped on the device to form an electric double layer capacitor between transistor channel and gate electrode. The ionic liquid was prebaked at 120 °C under the vacuum condition to remove moisture.

### 2. Microstructure characterization and property measurements

## 2.1 Microstructural characterization

The microstructure of nanoporous graphene samples was characterized by scanning electron microscope (SEM, JEOL JSM−6700) and transmission electron microscope (JEOL JEM−2100F) equipped with two aberration correctors for the image- and probe-forming lenses. SEM observations were conducted at an accelerating voltage of 15.0 kV and the samples were placed on conductive carbon tapes. For TEM observations, the samples were transferred on a Cu grid without a carbon support film. 3D imaging of nanoporous graphene was retrieved by FIB/SEM tomography **[5]** with a JEOL JSM−6700 multi-beam system operated at 15 kV for SEM and 30 kV for FIB slicing. The sample was repeatedly milled with the FIB and each newly produced block face was imaged by SEM for the acquisition of tomographic datasets. The slice and view process were repeated 130 times to obtain the dataset. The subsequent imaging process, including image stacking, alignment, binarization, segmentation and 3D imaging, was performed using a software package (TRI/3D, Ratoc System Engineering, Tokyo, Japan). Nanopore sizes and surface areas of the nanoporous graphene samples were measured with the Brunauer−Emmett−Teller (BET) method and Barrett−Joyner−Hallender (BJH) method using a BELSORP-mini II (BEL. JAPAN. INC) at 77 K. The horizontal axis was normalized with the vapor pressure of nitrogen ($P_0$) at 77.0 K (= 0.101 MPa). All samples were heated at 120°C under vacuum for 48 hours before the measurements. The mass of the samples was measured with an ultramicro balance.

## 2.2 Raman spectroscopy of nanoporous graphene

Raman spectra were recorded by using a micro-Raman spectrometer (Renishaw InVia RM 1000) with an incident wavelength of 514.5 nm. The laser power was set at 2.0 mW to avoid possible damage by laser irradiation. The nanoporous graphene samples were placed on a background-free glass slide. The accumulation time of each spectrum is 400 s.

## 2.3 Photoelectron emission spectroscopy of nanoporous graphene

Photoemission spectroscopy (PES) measurements were performed using a MBS A−1 spectrometer with a high-flux helium discharge lamp and a toroidal grating monochromator at room temperature. He IIα resonance line ($hv$ = 40.814 eV) was employed for photoelectron excitation. The energy resolutions were set at 32 meV. Before PES measurements, the nanoporous graphene samples were annealed at ~600°C under ultrahigh vacuum at $1\times10^{-10}$ Torr for 3 hours. The Fermi levels of the samples were referred to that of a gold film deposited on the sample holder.

## 2.4 Measurements of transport properties and capacitance.

The nanoporous graphene EDLT devices were measured using the semiconductor parameter analyzer (B1500, Agilent), signal generator (6221, Keithrey) and nanovoltmeter (34420A, Agilent) at 300 K and the magnetic fields ranging from −9 to 9 T. The capacitance was measured using the potentiostat/galvanostat (IVIUM Technologies) under a constant voltage between the transistor channel and the gate ($10^0$ - $10^5$ Hz). The total capacitance is described by: $C = \left(\frac{1}{C_Q} + \frac{1}{C_L}\right)^{-1}$, where $C_Q$ is a quantum capacitance and $C_L$ is a geometrical capacitance, respectively. For the EDLT devices, a huge geometrical capacitance is expected but a quantum capacitance, being proportional to finite density of states of graphene, is dominant in the total capacitance **[6-8]**. The capacitance of the transistor channel is calculated with the assumption that the surface area of the reference electrode is sufficiently large. In order to control the device temperatures and applied magnetic fields, the physical property measurement system (PPMS, Quantum design) was employed. The nanoporous graphene EDLT devices attached on the insertion probe of the cryostat were inserted to the sample space of PPMS. The sample space was evacuated using the cryopump at 360 K for 1 day to reduce the moister in the ion liquid.

### 3. Electric properties of nanoporous graphene.
### 3.1 Gate voltage dependence of electric conductance of nanoporous graphene EDLT devices.

The gate voltage ($V_G$) dependence of electric conductance ($\sigma$) was investigated for the nanoporous graphene EDLT device (**Fig. S2**). A small hysteresis observed from the nanoporous graphene devices is similar to the previous report in the EDLT devise of the carbon nanotube [**4**]. As shown in **Fig. S2 (a)** and **(b)**, a well-defined V−shaped $V_G$−$\sigma$ characteristic with a conductance minimum at around −0.6 V can be obtained from the nanoporous graphene device with the pore size of ~100 – 300 nm and ~ 1 μm. Note that for the ~ 1 μm pore sized graphene $\sigma$ tends to saturate at around $V_G$ = −1.6 and 0.4 V. In 2D graphene EDLT devices, the local saturation in $V_G$−$\sigma$ gradient is related to the effect of sub-bands in bilayer and trilayer graphene [**6**]. In fact, the Raman spectra as shown in **Table S1** demonstrate a smaller $I_{2D}/I_G$ of the 1μm pore sized nanoporous graphene than that of the 100 – 300 nm one [**2, 9, 10**]. Therefore, it appears that the 1 μm pore sized nanoporous graphene is composed of both monolayer and bilayer graphene sheets. By contrast, for the 100 – 300 nm pore nanoporous graphene device, $V_G$−$\sigma$ shows a distinct V−shape in the measured $V_G$ range, which is consistent with the monolayer graphene feature of the 100 – 300 nm pore nanoporous sample [**11**].

### 3.2 Semiclassical theory for the magnetotransport in nanoporous graphene

When the nanoporous graphene is placed in a uniform magnetic field, the relative angle between the magnetic field and the normal of the curved graphene surface varies from place to place. Since the electron orbital motion is only affected by the magnetic field component perpendicular to the plane, an electron traveling on the graphene labyrinth feels a non-uniform magnetic field, and thus the total conductivity should be expressed as an average of the 2D conductivity over the magnetic field amplitude when the inhomogeneity varies slowly. Here we present a semiclassical picture to describe the magnetotransport along this line and show that the approach well describes the magnetic-field dependence of the Hall resistance and the magnetoresistance observed in the experiments.

We first consider a flat two−dimensional electron system in a perpendicular magnetic field $B$. In the semiclassical approximation, the conductivity tensor is given as

$$\begin{pmatrix} \sigma_{xx}^{2D} & -\sigma_{xy}^{2D} \\ \sigma_{xy}^{2D} & \sigma_{xx}^{2D} \end{pmatrix} = \frac{\sigma_0^{2D}}{1 + \omega_c^2 \tau^2} \begin{pmatrix} 1 & -\omega_c \tau \\ \omega_c \tau & 1 \end{pmatrix},$$

where $\omega_c$ is the cyclotron frequency, $\tau$ is the scattering time, and $\sigma_0^{2D}$ is the conductivity in zero magnetic field. For graphene, the conductivity is simplified as [**11**]

$$\sigma_0^{2D} = n_s e \mu,$$

where $n_s$ is the carrier concentration and $\mu$ is the mobility given by

$$\mu = \frac{e v^2 \tau}{\varepsilon_F}.$$

here $v$ is the constant band velocity of graphene, and $\varepsilon_F$ is the Fermi energy. The cyclotron frequency is expressed as

$$\omega_c = \frac{e v^2 B}{\varepsilon_F},$$

so we end up with a relation

$$\omega_c \tau = \mu B,$$

as in ordinary metals.

Now let us consider the conductivity of 3D nanoporous graphene in a uniform magnetic field $B$ along $z$-direction. Here we assume that the normal of the local graphene surface is distributed isotropically, and that the conductivity contributed by surface elements are additive, although a more accurate picture would require a sophisticated method such as the random network theory. The total three-dimensional conductivity tensor then becomes

$$\sigma_{ij}(B) = \frac{S_{tot}}{V} \int_0^{\pi/2} \sigma_{ij}^{2D}(B\cos\theta) \sin\theta \, d\theta,$$

where $\theta$ is the angle from the surface normal to the magnetic field, $S_{tot}$ and $V$ are the total graphene area and the total volume of the system, respectively. The integral can be analytically evaluated to give

$$\sigma_{xx}(B) = \sigma_0 \frac{\tan^{-1}\omega_c\tau}{\omega_c\tau}, \qquad \sigma_{xy}(B) = \sigma_0 \frac{\log(1+\omega_c^2\tau^2)}{2\omega_c\tau},$$

where $\sigma_0 = (S_{tot}/V)\sigma_0^{2D}$. The resistivity tensor is given by inverting the conductivity tensor as

$$\rho_{xx}(B) = \rho_0 \omega_c \tau \frac{\tan^{-1}\omega_c\tau}{(\tan^{-1}\omega_c\tau)^2 + [\log(1+\omega_c^2\tau^2)]^2/4},$$

$$\rho_{xy}(B) = \rho_0 \omega_c \tau \frac{[\log(1+\omega_c^2\tau^2)]/2}{(\tan^{-1}\omega_c\tau)^2 + [\log(1+\omega_c^2\tau^2)]^2/4},$$

where $\rho_0 = [(S_{tot}/V)\sigma_0^{2D}]^{-1}$. In **Fig. 4 (c) and (d)** in the main text, we have plotted the magnetoresistivity $(\rho_{xx} - \rho_0)/\rho_0$, and the normalized Hall resistivity $\rho_{xy}/\rho_0$ against $\omega_c \tau = \mu B$. We have also plotted the second derivative of the Hall resistivity, $d^2\rho_{xy}/dB^2$, in **Fig. 4(d)**. The plots fully agree with the qualitative features in the experimental results, including the characteristic kink structure in $\rho_{xy}$ (i.e., the peak in the second derivative) appearing at $\mu B \sim 0.7$.

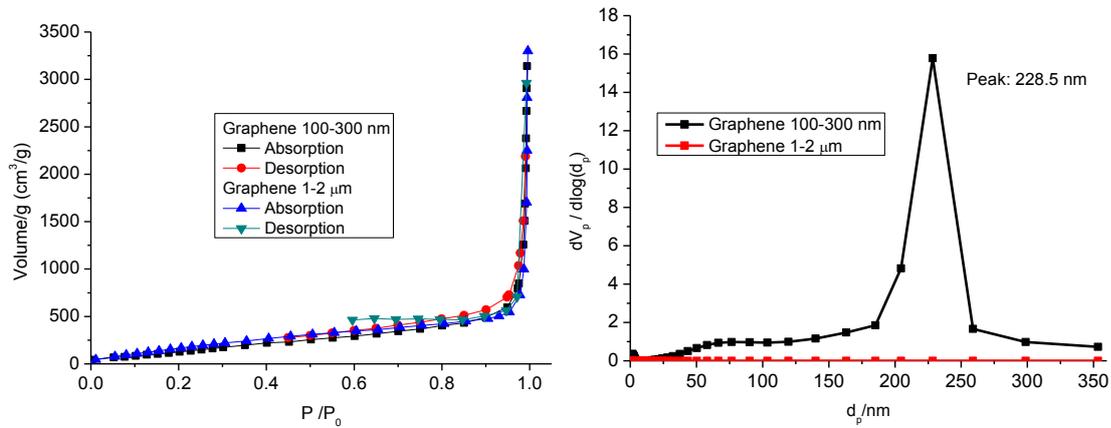

**Figure S1.** Brunauer–Emmett–Teller (BET) and Barrett-Joyner-Hallender (BJH) measurements of nanoporous graphene (100-300 nm). The peak pore diameter of 100-300 nm samples is 228 nm and the no peak pore diameter of 1-2 μm samples observed in the measureable range of 0-350 nm with this equipment.

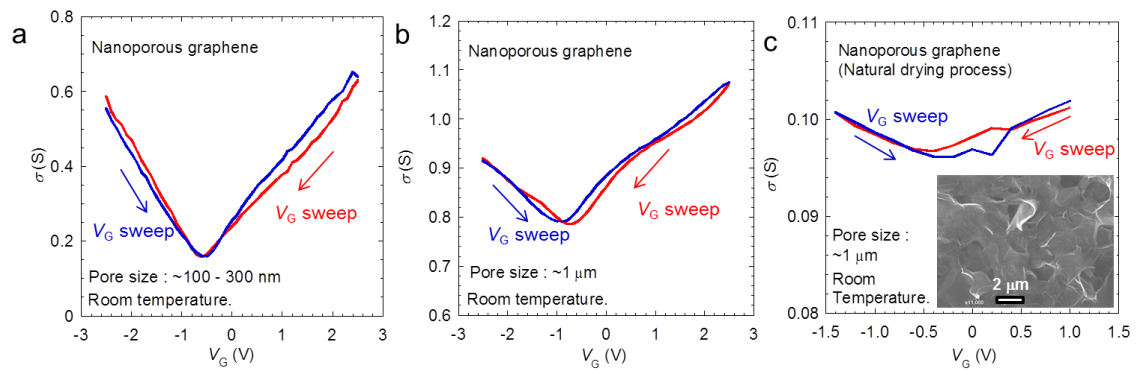

**Figure S2.** Gate voltage ($V_G$) dependence of electric conductance ($\sigma$) for nanoporous graphene (a) with ~ 100 – 300 nm pore; (b) with ~1 μm pore; and (c) with ~1 μm pore prepared by a natural drying process. Inset of (c) shows the SEM image of nanoporous graphene prepared by the natural drying process, in which collapsed pores can be seen. The collapse and damage of nanoporous structure lead to ~25 time smaller capacitance ($C$ ~ 0.04 mF/cm$^2$) and significantly small variation range of $\sigma$, in comparison with the nanoporous graphene prepared by supercritical drying. Consequently, the damage-free supercritical drying process is essential for achieving high device performances of nanoporous graphene.

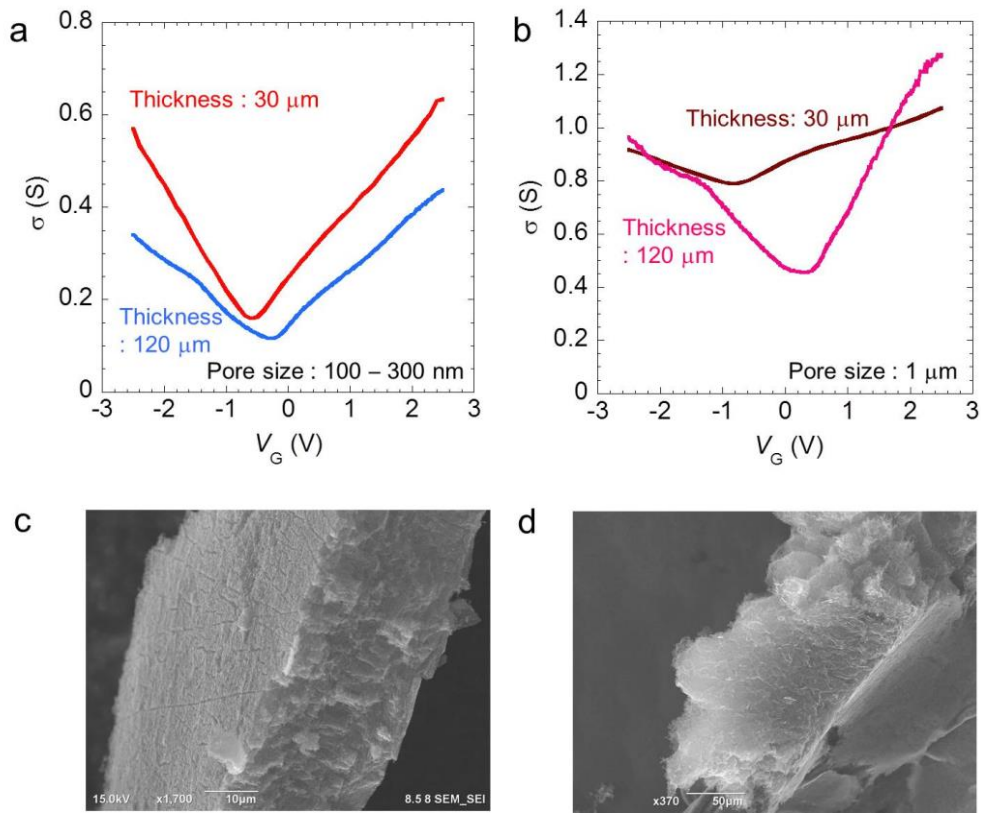

**Figure S3.** Thickness dependence of $\sigma - V_G$ curves for nanoporous graphene with ~ 100 – 300 nm pore (a) or with ~1 μm pore (b). (c,d) SEM image of nanoporous graphene (thickness of 120 μm) with ~ 100 – 300 nm pore (c) ~1 μm pore (d).

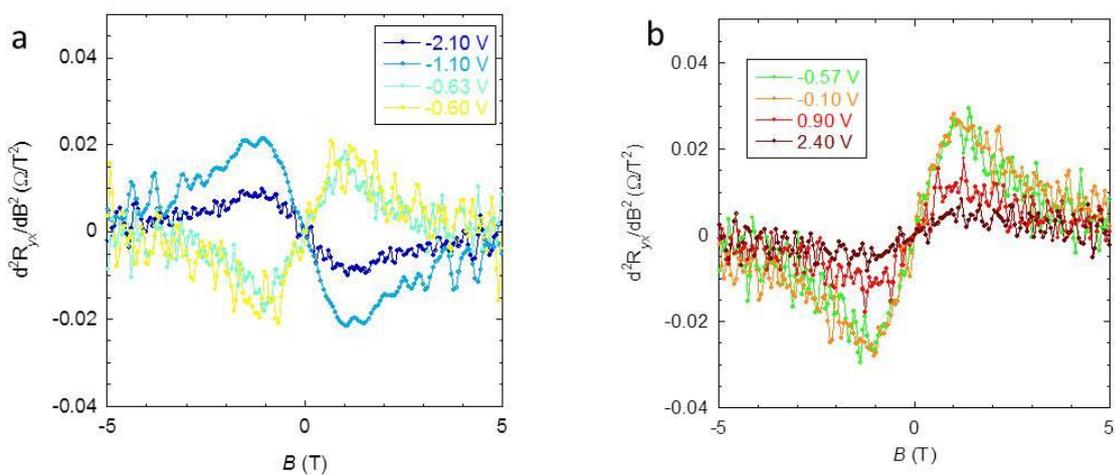

**Figure S4.** The second derivative of $R_{yx}$ ($d^2R_{yx}/dB^2$) for the nanoporous graphene at various $V_G$.

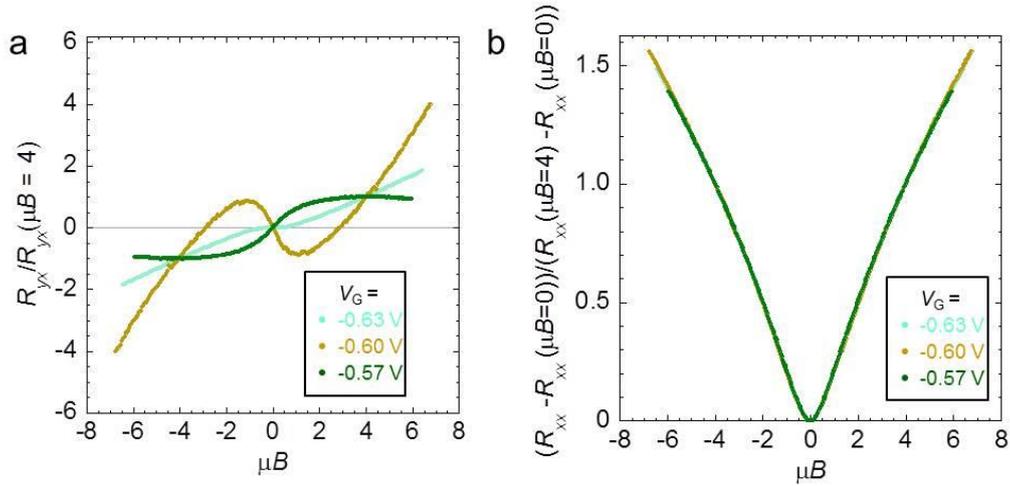

**Figure S5.** μ$B$ dependence of (a) $R_{yx}/R_{yx}(\mu B = 4)$; and (b) $[R_{xx}-R_{xx}(\mu B=0)]/[R_{xx}(\mu B = 4)-R_{xx}(\mu B=0)]$ for $V_G$ = -0.63, -0.60, -0.57 where the chemical potential is close to the Dirac point.

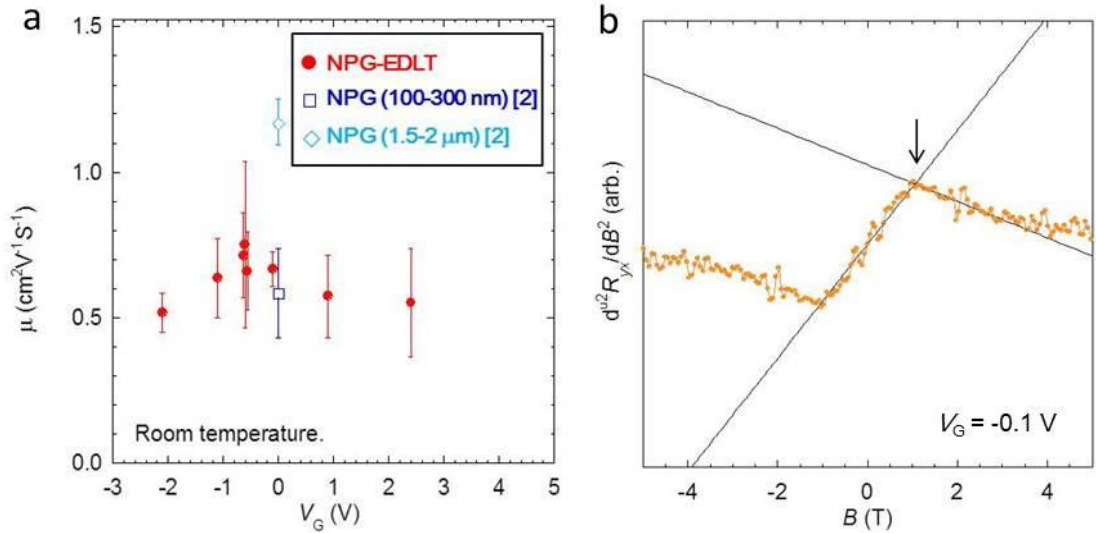

**Figure S6.** (a) $V_G$ dependence of the carrier mobility μ; and (b) $d^2R_{yx}/dB^2$-$B$ curves. In the semiclassical transport theory for the nanoporous graphene as described in the section 4.3, the $d^2R_{yx}/dB^2$ shows the peak (dip) at $\mu B \sim 0.7$. In the present study, μ values are evaluated using the peak (dip) position in $d^2R_{yx}/dB^2$-$B$ curves. As displayed in (b), $d^2R_{yx}/dB^2$-$B$ curvature around the peak was fitted by the linear function and the crossing point was employed as the peak position. The error bar was evaluated from the dispersion of data.

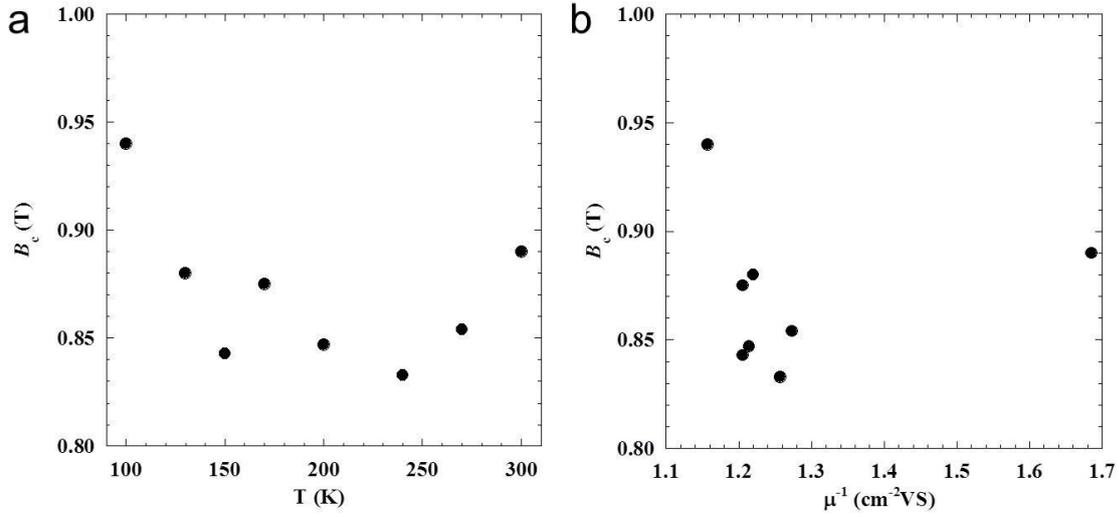

**Figure S7.** (a) Temperature dependence of the crossover magnetic field $B_c$ for the nanoporous graphene with a pore size of 100 – 300 nm. $B_c$ is the crossover magnetic field at which the magnetoresistance tends to change from quadratic to linear behaviour. (b) $B_c$ dependence of carrier mobility $\mu^{-1}$ evaluated by using the peak position in $d^2R_{yx}/dB^2$-$B$ curves. As pointed in the theoretical analysis [12], when the scattering from disorders or inhomogeneities of a system give rise to the linear magnetoresistance, the crossover magnetic field $B_c$ is derived as $B_c = \langle\mu\rangle^{-1}$ ($\mu$: carrier mobility) and expected to increase continuously with increasing temperatures. In our work, neither $B_c = \langle\mu\rangle^{-1}$ nor a continuous increase of $B_c$ against temperature can be observed. Apparently, the present magnetotransport behavior is not due to the effect of the inhomogeneity or disorder of the system.

**Table S1.** Raman band measurements of the nanoporous graphene used in this study. Unit of the spectra and line width is cm$^{-1}$.

|  | D band line width | G band line width | D' band line width | 2D band line width | $I_D/I_G$ | $I_{2D}/I_G$ |
|---|---|---|---|---|---|---|
| Graphene (pore size: 1.0 µm) | 1356<br>23 | 1583<br>11 | -<br>- | 2712<br>27 | - | 2.0 |
| Graphene (pore size: 100-300 nm) | 1358<br>23 | 1588<br>12 | -<br>- | 2707<br>27 | 0.16 | 2.4 |